\documentclass[
 reprint,
superscriptaddress,
 amsmath,amssymb,
 aps, prx, numbers, floatfix,longbibliography
]{revtex4-2}
\usepackage[symbol]{footmisc}
\usepackage{amssymb}
\usepackage{subcaption}
\usepackage{caption} % To left-align main caption
\usepackage{dsfont}
\usepackage[dvipsnames]{xcolor}
\usepackage{graphicx}
\usepackage{dcolumn}
\usepackage{bm}
\usepackage{braket}
\usepackage{placeins}
\usepackage{hyperref}
\begin{document}
\newcommand{\ckp}{\mathrm{C}_k\mathrm{P}}
\newcommand{\ckz}{\mathrm{C}_k\mathrm{Z}}

\title{Parametrized multiqubit gate design for neutral-atom based quantum platforms}
\author{Madhav Mohan}
\thanks{Send all correspondence regarding the article to \href{mailto:m.mohan@tue.nl}{m.mohan@tue.nl}}

\affiliation{Department of Applied Physics and Eindhoven Hendrik Casimir Institute, Eindhoven University of Technology, P.O. Box 513, 5600 MB Eindhoven, the Netherlands}
\author{Julius de Hond}
\affiliation{PASQAL SAS, Fred.\ Roeskestraat 100, 1076 ED Amsterdam, The Netherlands}
\author{Servaas Kokkelmans}
\affiliation{Department of Applied Physics and Eindhoven Hendrik Casimir Institute, Eindhoven University of Technology, P.O. Box 513, 5600 MB Eindhoven, the Netherlands}

\date{\today}
\preprint{APS/123-QED}

\begin{abstract}

A clever choice and design of gate sets can reduce the depth of a quantum circuit, and can improve the quality of the solution one obtains from a quantum algorithm. This is especially important for near-term quantum computers that suffer from various sources of error that propagate with the circuit depth. Parametrized gates in particular have found use in both near-term algorithms and circuit compilation. The one- and two-qubit versions of these gates have been demonstrated on various computing architectures. The neutral atom platform has the capability to implement native $N$-qubit gates (for $N \geq 2$). However, one needs to first find the control functions that implement these gates on the hardware. We study the numerical optimization of neural networks towards obtaining \emph{families} of controls -- laser pulses to excite an atom to Rydberg states -- that implement phase gates with one and two controls, the $\mathrm{C_1P}$ and $\mathrm{C_2P}$ gates respectively, on neutral atom hardware. The pulses we obtain have a duration significantly shorter than the loss time scale, set by decay from the Rydberg state. In addition, they do not require single-site addressability and are smooth. Hence, we expect our gates to have immediate benefits for quantum algorithms implemented on current neutral atom hardware. 
\end{abstract}
\maketitle

%\tableofcontents
\section{\label{sec:level1}Introduction}

Quantum devices based on neutral atoms offer a unique approach to quantum information and computing. In these systems, individual atoms can be trapped in optical traps, called \emph{tweezers}. Arrays with up to $6100$ atoms \cite{manetsch2024} have been constructed -- which demonstrates the promising scalability of the platform. Movement of trapped atoms in the array has also been demonstrated \cite{Bluvstein_2022}. This is termed dynamic connectivity, and can be used to swap or entangle distant qubits in a grid. These features, combined with demonstrated two-qubit gate fidelities of $F = 0.997$ \cite{tsai2024,universal2024}, and error suppression using error correcting codes \cite{Bluvstein_2023} make the platform competitive with other leading platforms such as superconducting circuits and trapped ion devices.

\par In neutral-atom architectures, quantum information is generally encoded in the low-lying electronic states of single atoms. Entanglement is mediated by exciting the atoms into high-lying electronic states -- that is, states with a large principal quantum number $n$, termed \emph{Rydberg states} -- through lasers resonant with the desired transition. The interactions between such states are strong and long-ranged, and hence allow for the implementation of native multiqubit gates -- that is, $N$-qubit gates with $N \geq 2$. Such gates have wide ranging applications in quantum error correction \cite{PhysRevA.101.022308, PhysRevLett.117.130503, PhysRevA.108.062426}, quantum algorithms \cite{M_lmer_2011, de_Keijzer_2022} and quantum compilation \cite{patel2022, staudacher2024}. 
\par
To implement these gates on hardware, however, one must devise pulses -- that is, the time profile of the laser intensity and frequency -- that can carry out the desired unitary transformation on the electronic states of the atoms. On the theoretical side, such pulses have been proposed for $k$-qubit Toffoli and fan-out gates (for $k>2$)~\cite{Jandura_2022, PhysRevX.10.021054, Evered_2023, isenhower2011, PhysRevA.101.022308, yu2022}, as well as for $k$-controlled $\mathrm{SWAP}$ gates~\cite{wu2021, yang2024} and more general unitaries~\cite{song2024}. Such gates have also been demonstrated experimentally on the Rydberg platform \cite{Evered_2023, cao2024multiqubit, PhysRevLett.123.170503}. Notably, Cao \emph{et al.}\ implemented $N$-qubit rotation gates (for $N$ up to 9) in order to generate cascaded GHZ states \cite{cao2024multiqubit} by placing $N$ atoms within interacting range of each other. The pulses devised to implement this gate are also \emph{time-optimal} -- meaning they are the shortest possible pulses that implement the corresponding unitary. Such pulses that implement the gate directly are significantly shorter in duration than implementing the decomposition of the gate in terms of one- and two-qubit gates \cite{PhysRevLett.129.050507}, and are obtained numerically using a class of methods termed \emph{quantum optimal control} (QOC) \cite{Jandura_2022, koch2022}. 
\begin{figure*}[htbp]
\centering
\captionsetup{justification=raggedright,singlelinecheck=false}

\includegraphics[width=0.9\textwidth]{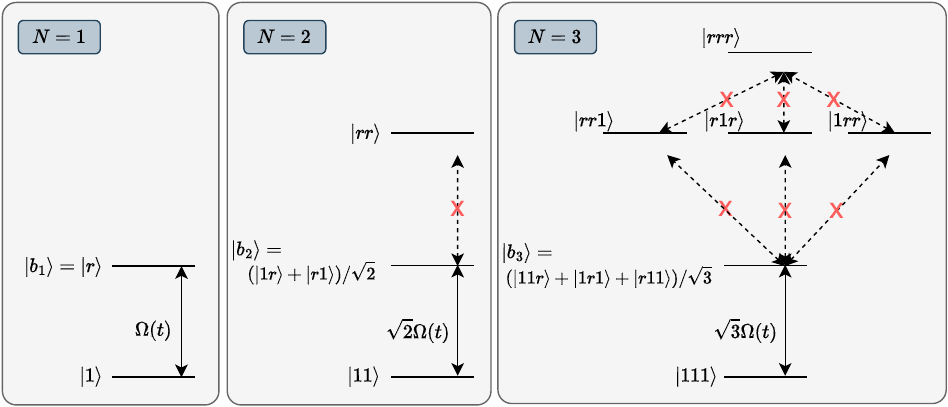}

\caption{Relevant level diagrams for the $\ckp$ gate pulses. Of the computational basis states, only $\ket{1}$ is drawn since $\ket{0}$ is not coupled to $\ket{r}$. The left, middle, and right panels correspond to the one, two, and three-qubit case, respectively. Not all the levels involved are illustrated -- for instance, $\ket{110}$ can be understood in terms of the $\ket{11}$ level for the two-qubit case. Due to the Rydberg interaction, higher excited states are detuned and are not coupled, as indicated by the red crosses.}
\label{fig:tls}
\end{figure*}

In particular, we highlight the variational neural network (NN) methods that have gained prominence in quantum control problems~\cite{zheng2021,PhysRevX.8.031084,niu2019,dalgaard2020,PhysRevResearch.4.L012029, PhysRevLett.129.050507}. These networks are variational ansatze with a structure inspired by the networks of neurons used for learning in the brain. Importantly, they are universal approximators of multivariate, continuous functions~\cite{Kurt1991251}, and the exact gradients over their weights can be computed numerically~\cite{margossian2019}. This means that NN-based methods can generalize to challenging QOC tasks, including control of driven systems~\cite{PhysRevResearch.4.L012029} and systems including quantum feedback~\cite{PhysRevX.8.031084}.

Recent studies have also investigated the parametrized versions of multiqubit gates -- gates that implement some unitary $U(\bm{\phi})$, for $M$ parameters \mbox{$\bm{\phi} = (\phi_1, \phi_2, \dots, \phi_M)$}.  In terms of variational algorithms, there are proposed applications of such gates towards simulating lattice gauge theories \cite{PhysRevX.13.031008} and combinatorial optimization \cite{Dlaska_2022}. Further, for a given circuit, the inclusion of $k$-controlled phase, or $\ckp$ gates (for $k \geq$ 2) in the gate set can reduce the gate count and the total circuit evaluation time of the compiled circuit \cite{ staudacher2024}, thus reducing infidelities -- as most error sources propagate in time \cite{Jandura_2022}. 
% now for what we do 
This reduction is especially significant for the current era of quantum hardware, termed the noisy intermediate-scale quantum (NISQ) era. In these devices, the coherence times of the qubits are limited, and there are multiple sources of gate infidelity. 

We note that preliminary work exists on $\ckp$ protocols for neutral atoms, in particular for the $\mathrm{C_1P}$~\cite{fromonteil2024, Evered_2023} and $\mathrm{C_2P}$~\cite{PhysRevResearch.7.013034} gates. A common limitation across these studies is the necessity of re-optimizing control parameters for each specific gate angle $\phi$. Such protocols return control function(s) that implement the gate at a fixed $\phi$ -- whereas the complete protocol must be a family of controls, that is, continuous (or piecewise continuous) in $\phi$~\cite{PhysRevLett.129.050507}. We will discuss these studies further in Sec. \ref{multiq}. 

We investigate the problem of finding families of pulses that implement the $\mathrm{C_1P}$ and $\mathrm{C_2P}$ gates on Rydberg hardware. Employing numerical techniques from QOC, we find pulses for both gates that are fast -- more specifically, the $\mathrm{C_1P}$ pulse we obtain is time-optimal, and the $\mathrm{C_2P}$ pulse operates on a time scale comparable to the time-optimal solution. These pulses are also smooth, and require only \emph{global} addressing -- the same pulse is sent to all atoms -- and hence can be implemented on NISQ devices.

To the best of our knowledge, our paper hence presents the first $\ckp$ pulse families on the Rydberg platform. Further, the techniques we use for the optimization can be extended to gates over more control qubits, as well as to fast quantum state preparation for parametrized states.

This work is structured as follows. In Section \ref{theory} we discuss the physics of our system -- including the blockade mechanism by which entanglement is generated, as well as the dynamics of the quantum states when pulses are applied to the atoms. We also expand upon the optimization techniques we use to obtain the pulses. In Section \ref{sec:results} we present the results of the numerical optimization -- the pulses we obtain -- and study features of interest, including the gate time and sources of error. Finally, in Section \ref{sec:discussion} we conclude by discussing future directions of research.

\section{Theory}
\label{theory}
\subsection{Multiqubit gates}
\label{multiq}

In the neutral atom architecture, we consider the \emph{ground-ground} (gg) qubit scheme \cite{Morgado_2021} -- where quantum information is encoded in low-lying electronic states of neutral atoms. Such states are long lived and can allow for excellent coherence times of the qubits \cite{Madjarov_2020}. As an example we consider the strontium-$88$ ($^{88}\mathrm{Sr}$) atom, but note that the protocols we will obtain also work for other atoms -- popular choices being ytterbium-$141$ \cite{thompson} and rubidium-$87$ \cite{Evered_2023}. 

We encode $\ket{0}$ in the ground $5\mathrm{s}^2 \,^1\mathrm{S}_0$ state and $\ket{1}$ in the metastable $5\mathrm{s}5\mathrm{p} \,^3\mathrm{P}_0$ state. The Rydberg state $\ket{r}$ should have a large enough $n$ such that strong interactions are generated -- for instance, the $5\mathrm{s}61\mathrm{s}\,^3\mathrm{S}_1$ state was chosen by Tsai \emph{et al.} in their demonstration of the current platform-leading two-qubit gate fidelities \cite{tsai2024}. 

The problem we are concerned with is \emph{family control} \cite{PhysRevLett.129.050507} -- given a target unitary transformation $U_{\mathrm{tgt}}(\bm{\phi})$ for parameters $\bm{\phi} = (\phi_1, \phi_2, \dots, \phi_M)$, obtain the \emph{family} of control pulses $\mathbf{C}(\bm{\phi}, t)$ that implement the transformation for some control hamiltonian $H_C$, on quantum hardware. By family of pulses, we hence mean control functions that are continuous in $\bm{\phi}$. For a neutral atom device with $N$ atoms, this family of controls includes the Rabi frequency 
\begin{equation}
\bm{\Omega}(\bm{\phi}, t) = (\Omega_1(\bm{\phi}, t), \Omega_2(\bm{\phi}, t),\dots,\Omega_N(\bm{\phi}, t)),
\end{equation}
 and the detuning 
\begin{equation}
\bm{\Delta}(\bm{\phi}, t) = (\Delta_1(\bm{\phi}, t), \Delta_2(\bm{\phi}, t),\dots,\Delta_N(\bm{\phi}, t)),
\end{equation} 
that are proportional to the laser intensity and laser frequency respectively. We note that the controls are vectors in the general case, as one could consider addressing each atom individually with a separate laser. Here $\Omega_i(t) \propto \sqrt{P_i}$ -- for $P_i$ the power of the $i^{\mathrm{th}}$ laser --  and $\Delta_i(t) = (\omega_{L,i}(t) - \omega_{1,r})$, for $\omega_{L,i}(t)$ the frequency of the $i^{\mathrm{th}}$ laser. Hence, 
\begin{equation}
\label{ctrl}
\mathbf{C}(\bm{\phi}, t) = (\bm{\Omega}(\bm{\phi}, t), \bm{\Delta}(\bm{\phi}, t)).
\end{equation}

We aim to find hardware controls to implement, natively, the $k$-controlled phase gates \cite{staudacher2024} for $k=1, 2$ -- corresponding to a two-qubit and three-qubit gate respectively. 

\begin{figure}
\hspace*{0cm}
\captionsetup{justification=raggedright,singlelinecheck=false}
\includegraphics[width = 0.4\textwidth]{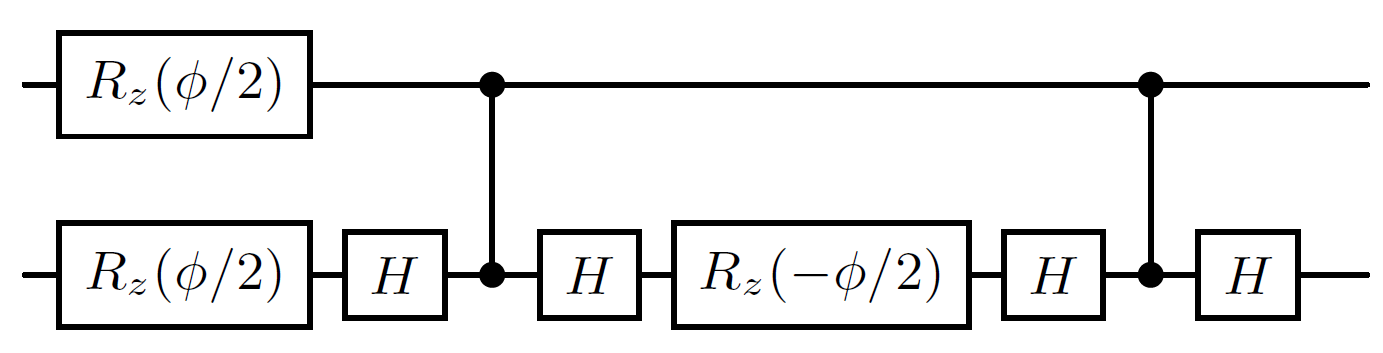}
\caption{The decomposition of a $\mathrm{C_1P(\phi)}$ gate into the standard gate set of a Rydberg device requires two $\mathrm{C_1Z}$ gates and several single-qubit gates. It has a longer pulse duration, and hence more error-prone, than a \emph{native} implementation of the $\mathrm{C_1P}$ gate which we present in Sec. \ref{sec:c1pr}.}
\label{fig:native}
\end{figure}

The corresponding target unitaries are 
\begin{equation}
\label{uc1p}
U_{\mathrm{C_1P}}(\phi) = \ket{0}\bra{0}_1\otimes(\mathds{1}^{1})_{2} + \ket{1}\bra{1}\otimes \mathrm{P}(\phi)_2
\end{equation}
\begin{equation}
\label{uc2p}
U_{\mathrm{C_2P}}(\phi) = (\mathds{1}^{2} - \ket{11}\bra{11})_{1, 2}\otimes(\mathds{1}^{1})_3 + \ket{11}\bra{11}_{1,2}\otimes \mathrm{P}(\phi)_3,
\end{equation}
where the superscript on the identity $\mathds{1}$ denotes the number of qubits it acts on, the subscript denotes the label of the qubit(s), and $\mathrm{P}$ is the \emph{phase} gate
\begin{equation}
\mathrm{P}(\phi) = 
\begin{bmatrix}
1 & 0\\
0 & e^{i\phi}\\
\end{bmatrix}
.
\end{equation}
Multiqubit gates, including $\mathrm{C_2P}$ gates, have been realized natively on various platforms. In Appendix~\ref{appendix:decomp} we present an overview of platform-specific methods towards generating these gates.
One can always implement such transformations by decomposing them into a product of one- and two-qubit gates available on the hardware -- this is visualized for the $\mathrm{C_1P}$ gate in Fig. \ref{fig:native}. However, this approach has a significantly longer gate duration \cite{PhysRevLett.129.050507, PRXQuantum.2.010101} than the control-level approach we adopt, and we also present an in-depth comparison of these two approaches for our problem in Appendix~\ref{appendix:decomp}.

We note that as $\mathrm{P(\pi) = Z}$, $\ckp(\pi)$ is equivalent to a $\ckz$ gate. Hence, the problem we consider is the generalized version of the problem of obtaining a $\ckz$ pulse \cite{Jandura_2022, PhysRevResearch.4.033019, Mohan_2023, fromonteil2024, giudici2024}. 

To illustrate the the physics of the blockade gate, consider first the case where one applies a laser field near-resonant to the ground to Rydberg state transition (specifically, from $\ket{1}$ to $\ket{r}$), to a pair of atoms.
The single-qubit Hamiltonian, then, is 
\begin{equation}
\label{eq:1qham}
H_{1q}(\Omega(t), \Delta(t)) = \frac{\Omega(t)}{2}(\ket{1}\bra{r} + h.c.) - \Delta(t)\ket{r}\bra{r}.
\end{equation}

Note that we set $\hbar = 1$ throughout this article.  

As the time-optimal $\ckp$ pulses have constant laser intensity \cite{fromonteil2024}, we keep the Rabi frequency fixed to a constant in our optimization, $\Omega(t) = \Omega_{\mathrm{max}}$. 
We also enforce global pulses, that is, the same pulse is simultaneously applied to all atoms, and hence $\Delta_i(t) = \Delta(t)$. This simplifies the experiment considerably as only one laser is required to implement the gate, further, single site addressability -- an experimental challenge \cite{PhysRevA.101.062309, Jandura_2022} -- is not needed. The control function of Eq.~(\ref{ctrl}) hence reduces to $C_{\ckp}(\phi, t) = \Delta(\phi, t)$.
 
For $N$ atoms then, the physics is captured by the Hamiltonian
\begin{equation}
\label{1rtrans}
H_{Nq} = \sum_{i=1}^N H_{1q}^i(\Omega(t),\Delta(t))\Big |_{\Omega(t) = \Omega_{\mathrm{max}}} + \sum_{i=1}^N\sum_{j>i}^N H_{\mathrm{int}}^{i,j},
\end{equation}
where 
\begin{equation}
H_{1q}^i = \mathds{1}_1\otimes\mathds{1}_2\otimes\dots\otimes(H_{1q})_i\otimes\dots\mathds{1}_N,
\end{equation}
and, defining the operator
\begin{equation}
\hat{O}_k^{i, j} =
\begin{cases}
|r\rangle\langle r|, & \text{if } k \in \{i, j\}, \\
\mathds{1}, & \text{otherwise.}
\end{cases}
\end{equation}
the interaction term is
{\small
\begin{equation}
H_{\mathrm{int}}^{i,j} = -\frac{C_6}{|\bm{R}_i - \bm{R}_j|^6} \bigotimes_{k=1}^N \hat{O}_k^{i, j}.
\label{eq:interaction}
\end{equation}
}

Note that the indices $i,j$ denote the atom(s) the operator acts on. Here, $H_{\mathrm{int}}^{i,j}$ describes the van der Waals interaction between two atoms excited to the Rydberg state, with $C_6$ the van der Waals interaction coefficient and $\bm{R}_k$ the displacement vector of the $k^{\mathrm{th}}$ atom. 
We will only investigate systems with atoms placed equidistant from each other. Defining the pair-interaction energy as $V_{\mathrm{int}}^{i,j} = -C_6/|\bm{R}_i-\bm{R}_j|^6$, this implies that
\begin{equation}
V_{\mathrm{int}}^{i,j} = V, \,\,\,\,\forall\, (i,j) \in A\times A
\end{equation}
for $A = \{1,2,...,N\}$.
We will refer to $B = V/\Omega_{\mathrm{max}}$ as the \emph{blockade strength}. In this work we will assume that \mbox{$B \gg 1$} -- this is called the blockade regime, as the presence of one atom in the Rydberg state suppresses the Rydberg population of neigboring atoms \cite{Jaksch_2000}. 

\begin{figure}%[htbp]
    \centering
    \includegraphics[width=0.4\textwidth]{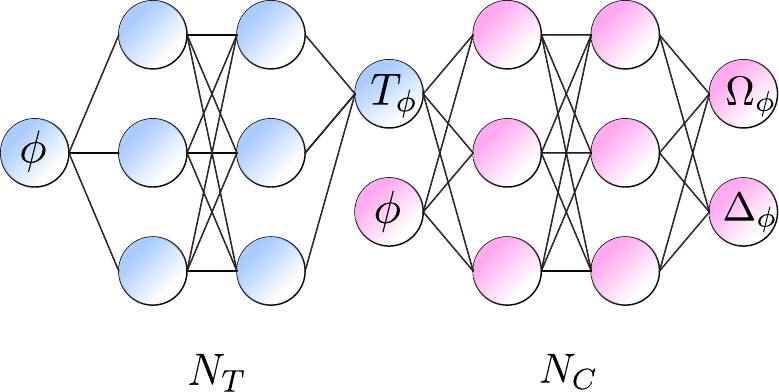}
    \captionsetup{justification=raggedright,singlelinecheck=false} 
    \caption{We visualize two \emph{chained} NNs, and label the input and output layers with the corresponding physical values. The intermediate layers consist of neurons with each edge representing a trainable weight in the weight matrices, $\bm{W}_T$ and $\bm{W}_C$ for $N_T$ and $N_C$ respectively. The neuron applies a weighted sum followed by an \emph{activation function} -- See Appendix \ref{nnarch} for details. This is an example of $N_{TC}$ -- the networks we use to find $C_{\ckp}(\phi, t)$ have $\Omega_{\phi}$ fixed to a constant.}
    \label{fig:network}
\end{figure}

For an $N$-qubit system in the blockade regime, evolving under the Hamiltonian from Eq.~(\ref{1rtrans}), one can realize unitaries that map each computational basis state to itself up to a phase \cite{fromonteil2024}. This means that the system is well suited to realizing $\mathrm{C}_{N-1}\mathrm{P}$ gates \cite{staudacher2024} -- upto a global single qubit $\mathrm{R_Z}$ rotation. The dynamics in the blockade regime can be understood in terms of two-level systems (TLS): two and three for the $\mathrm{C_1P}$ and $\mathrm{C_2P}$ gate, respectively. We illustrate these systems in Fig.~\ref{fig:tls}. The TLS are closed systems -- independent of each other, they partition the Hilbert space into subspaces with populations that do not mix \cite{fromonteil2024}. 

We consider first the $\mathrm{C_1P}$ gate, that acts on the computational space $\{\ket{00}, \ket{01}, \ket{10}, \ket{11}\}$, and is described by the matrix \mbox{$U = \mathrm{diag}(1,1,1,e^{i\phi})$} -- this is equivalent to the unitary from Eq.~(\ref{uc1p}). We note that $H_{2q}$ does not act on $\ket{00}$, as $H_{1q}$ does not act on $\ket{0}$. This implies that the dynamics of states $\ket{01}$ and $\ket{10}$ is equivalent, involving a transition $\ket{1}\leftrightarrow\ket{r}$ with coupling strength $\Omega_{\mathrm{max}}$, forming the first TLS. The transfer from the $\ket{11}$ state to the $\ket{rr}$ state is forbidden for the case of a perfect blockade ($B \to \infty$) and this state forms the second TLS along with the two-qubit \emph{bright} state 

\begin{equation}
\label{eq:2qb}
\ket{b_2} = \frac{1}{\sqrt{2}}(\ket{1r}+\ket{r1}),
\end{equation}
which involves the transition $\ket{11}\leftrightarrow\ket{b_2}$ with Rabi frequency $\sqrt{2}\Omega_{\mathrm{max}}$ \cite{fromonteil2024}. 
\par 

The $\mathrm{C_2P}$ gate is described by the matrix \mbox{$U = \mathds{1}^{3} - 2e^{i\phi}\ket{111}\bra{111}$} -- equivalent to the definition in Eq.~(\ref{uc2p}). The dynamics involves the two TLS described above, along with a third TLS involving the transition $\ket{111}\leftrightarrow\ket{b_3}$ with the Rabi frequency $\sqrt{3}\Omega_{\mathrm{max}}$, where 
\begin{equation}
\ket{b_3} = \frac{1}{\sqrt{3}}(\ket{11r}+\ket{1r1}+\ket{r11})
\end{equation}
is the three-qubit bright state. 

\par 
Assuming perfect blockade $B\to\infty$, we can hence project out the forbidden states with more than one Rydberg excitation \cite{Jandura_2022}. This leads to an effective Hamiltonian $H_{\mathrm{eff}}$, defined in Eq. \ref{eq:h_eff} for a three-qubit system.  

We note, however, that the blockade strength is finite in the experiment [as computed in Eq.~(\ref{eq:interaction})], and this is a potential source of infidelity for gates optimized at $B\to\infty$. For the $\mathrm{C_1P}$ gate, we will hence simulate the dynamics (during the optimization) under the full Hamiltonian from Eq.~(\ref{1rtrans}) with $B = 21.1$, an experimentally realistic value~\cite{PhysRevResearch.4.033019, Evered_2023} -- as opposed to simulating $H_{\mathrm{eff}}$. This is done in order to mitigate the finite blockade errors \cite{Mohan_2023}.

Motivated by reducing the computational expense for the higher-dimensional problem of the $\mathrm{C_2P}$ gate, we started by simulating the dynamics under $H_{\mathrm{eff}}$ instead. However, as we explore in Sec. \ref{sec:results}, this leads to pulses with high infidelities for finite $B$ -- and we used these solutions (optimized initially for $H_{\mathrm{eff}}$) as initial guesses for a subsequent finite-blockade optimization.  

When reporting the final fidelities in Sec. \ref{sec:results}, one must also consider the losses arising from spontaneous emission and blackbody radiation (BBR) from the Rydberg state~\cite{PhysRevA.72.022347}  at $300\mathrm{K}$. To simulate this effect, we add a non-Hermitian term \cite{PhysRevResearch.4.033019, Mohan_2023} 
\begin{equation}
H_{\mathrm{decay}} = i\frac{\Gamma_{n = 61}}{2}\ket{r}\bra{r},
\end{equation}
to $H_{1q}$ in Eq.~(\ref{eq:1qham}), which assumes that all decay occurs to states outside of the computational subspace -- an assumption justified for the system we studied in this article \cite{PhysRevResearch.4.033019, madjarov_entangling_2021}.
Here $\Gamma_{n=61}$ is chosen as the decay width of the $5\mathrm{s}61\mathrm{s}\,^3\mathrm{S}_1$ state in Sr, corresponding to a lifetime $1/\Gamma_{n=61} = 96.5\mathrm{\mu s}$ \cite{Mohan_2023}. This term is subsequently used to construct the full Hamiltonian $H_{Nq,\Gamma}$, in a manner similar to Eq.~(\ref{1rtrans}). 

\subsection{Numerical optimization}
\label{subsec:numop}

Using established QOC techniques such as Gradient Ascent Pulse Engineering (GRAPE) \cite{khaneja_optimal_2005, Jandura_2022} or semi-analytical approaches \cite{Evered_2023, fromonteil2024}, one can obtain a pulse to implement the $\ckp$ gate for some fixed angle $\phi$. 
Consider the use of $\ckp$ gates in variational algorithms, where one might run the GRAPE algorithm to obtain the gate pulses, before carrying out the quantum circuit -- such optimizations would need to be run after every iteration of the algorithm, as the gate angle $\phi$ is updated by a classical solver in each iteration \cite{Tilly_2022}. This can lead to a significant overhead, considering that the runtime for GRAPE is longer than the runtime of the quantum circuit.

Sauvage \emph{et al.}\ investigated the generation of parametrized gates by training neural networks -- used as \emph{variational ansatze} -- to model families of pulses \cite{PhysRevLett.129.050507}. We build upon their approach, and adapt it to our neutral atom system. As we will motivate in this section, compared to GRAPE, the NN only has to be trained once, and then outputs a high-fidelity pulse for any angle in the domain of the gate without requiring further optimizations.
\begin{figure*}[htbp]
    \centering
    % First image
    \begin{subfigure}[b]{0.485\textwidth}
        \centering
        \includegraphics[width=\textwidth]{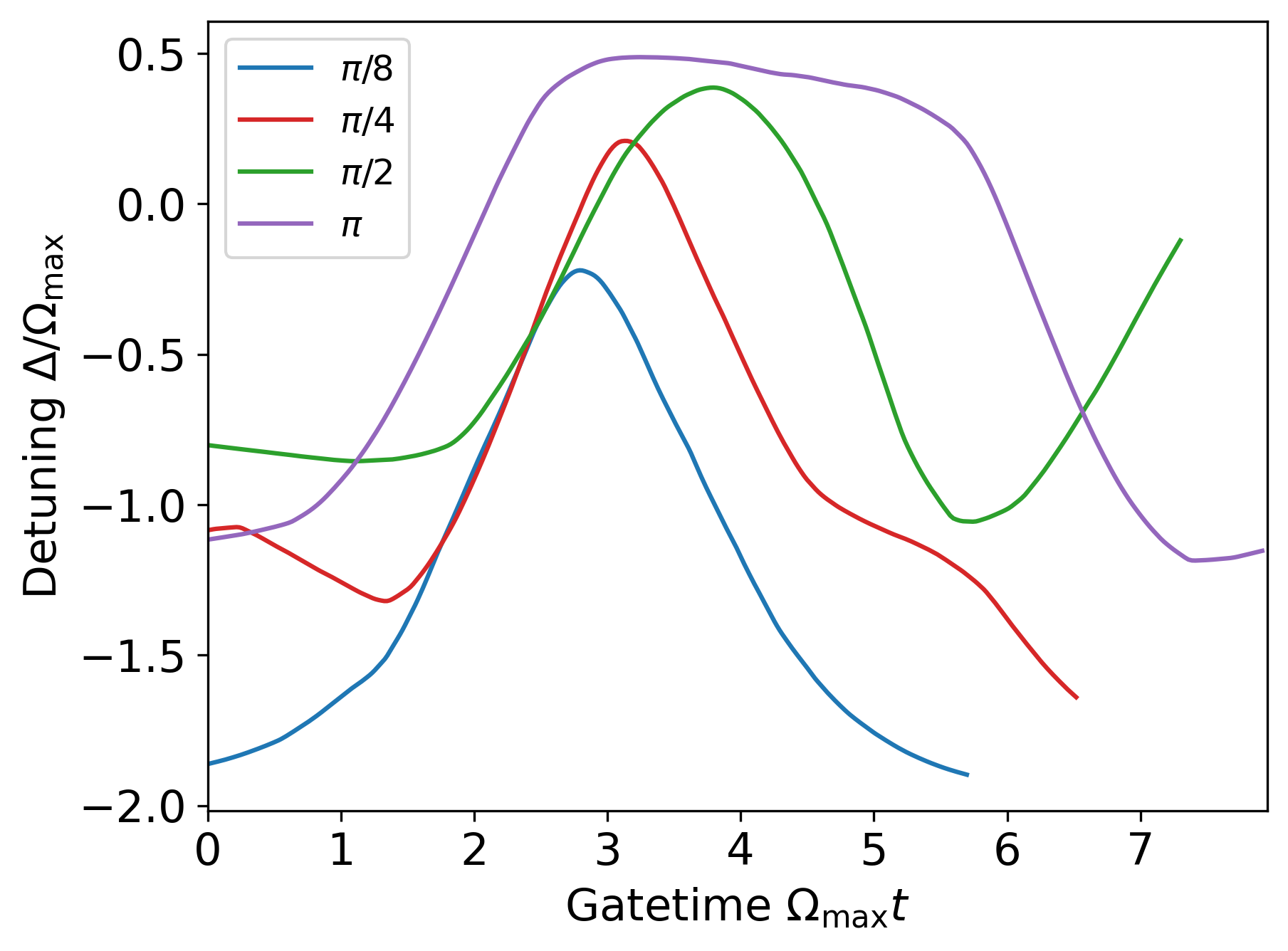}
        \caption{}
        \label{fig:pulses}
    \end{subfigure}
    \hfill
    % Second image
    \begin{subfigure}[b]{0.47\textwidth}
        \centering
        \includegraphics[width=\textwidth]{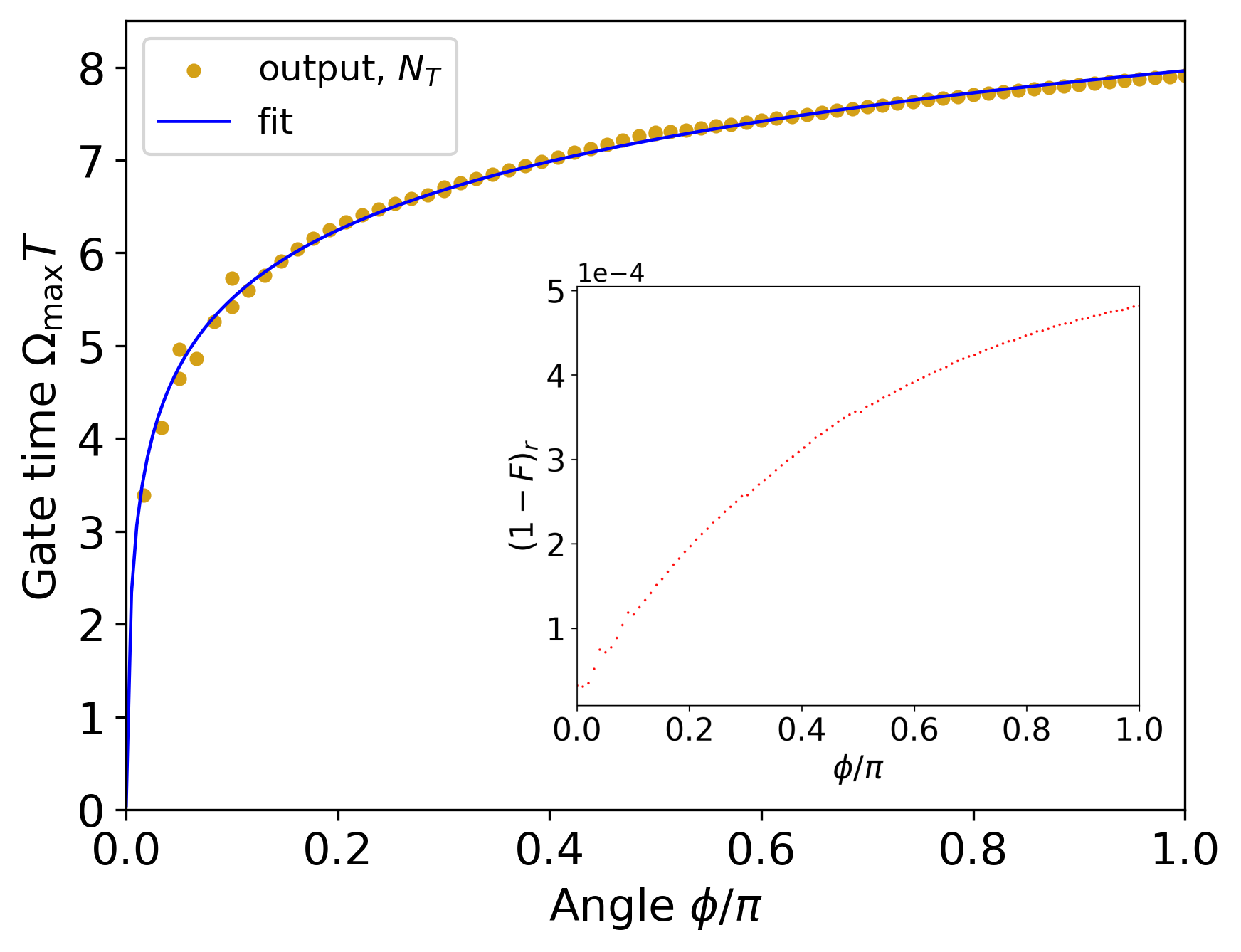}
        \caption{}
        \label{fig:timec1p}
    \end{subfigure}
    \captionsetup{justification=raggedright,singlelinecheck=false} % Only apply to this caption
    
    \caption{(a) Laser detuning $\Delta$ that implements, for different angles $\phi \in (0,\pi]$, the time-optimal $\mathrm{C_1P}$ gate. The laser intensity is kept fixed at the maximum value for the time-optimal pulses, such that the Rabi frequency $\Omega = \Omega_{\mathrm{max}}$. (b) The pulse time as a function of $\phi$  [NN output, fit to Eq.~(\ref{fitfn})]. Inset: We plot $(1-F)_r$ -- the infidelity arising due to decay from $\ket{r}$, as a function of $\phi$.}
    \label{fig:uni}
\end{figure*}

NNs consist of multiple layers, each with a collection of nodes, called \emph{neurons}, and trainable weights. In Appendix \ref{nnarch}, we discuss how these layers and neurons transform the inputs (in the form of vectors). NNs with the appropriate weights can approximate any continuous, multivariate function \cite{Kurt1991251}. 

The inputs to the NN -- which we call $N_C$ -- are $\phi$ and total pulse duration ${T}_{\phi}$. 
The angle $\phi$ is sampled randomly from the domain, $\mathrm{dom}(\ckp): (0,\pi]$. The open interval is chosen as, for $\phi = 0$,  $\ckp(0) = \mathds{1}$ is equivalent to not implementing a pulse, and $T_{\phi=0} = 0$. 
We request $C_{\ckp}(\phi,t)$ from the NN as the output. Once the pulses are obtained from the NN, we evolve the system under the Hamiltonian in Eq.~(\ref{1rtrans}) to obtain the output unitary
\begin{equation}
\label{time_evol}
U_{\mathrm{out}, k}({\phi}) = \mathcal{T} \mathrm{exp}  \left(-i\int_0^{T_{\phi}} H_{(k+1)q}(C_{\ckp}(\phi, t))dt\right),
\end{equation}
for $\mathcal{T}$ the time-ordering operator. Defining the projection operator $P$ onto the computational subspace~\cite{fidelity2007},

\begin{equation}
P = \sum_{i_1,i_2,\dots i_N\in\{0,1\}}\ket{i_1i_2\dots i_N}\bra{i_1i_2\dots i_N},
\end{equation}
the cost function at each iteration is 
\begin{equation}
\label{cost_fn}
J = 1 - \frac{1}{4^{(k+1)}}\braket{||U_{\ckp}(\phi)^{\dag}(PU_{\mathrm{out}, k}(\phi)P)||^2}_{\phi},
\end{equation}
where $||.||$ is the Hilbert-Schmidt norm and $\braket{.}_{\phi}$ implies we compute the mean over samples of $\phi$ drawn randomly from $\mathrm{dom}(\ckp)$ in each iteration. We will refer to $J = \braket{(1 - F)}_\phi$ as the mean \emph{infidelity} over the domain of the protocol. We then train the weights of the NN to minimize the cost function.

We first look at how the optimization works when the total pulse duration is fixed to some constant for each angle, $T_{\phi} = T$:
\begin{enumerate}
 
\item Initialize the weights $\bm{W}_C$ of $N_C$ randomly.
\item Construct a vector $\bm{v}_{\mathrm{in}}$ with $M$ angles drawn randomly from $\mathrm{dom}(\ckp)$, we fixed $M = 80$. Input $\bm{v}_{\mathrm{in}}$ to the network $N_C$. \item The network outputs $\bm{C}_{\ckp}(\bm{v}_{\mathrm{in}}, t)$ by carrying out transformations on $\bm{v}_{\mathrm{in}}$ as in Eq.~(\ref{forward}). We get a vector of control pulses as the output, one for each angle in $\bm{v}_{\mathrm{in}}$.
\item Compute $J$ by evolving the system as in Eq.~(\ref{time_evol}) with $T_{\phi} = T$, and computing the distance between the output unitary and the target unitaries from Eq.~(\ref{uc1p}) and~(\ref{uc2p}). 
\item Compute the gradients $\nabla_{\bm{W}_C} J$ i.e.\ the gradient of the cost function with respect to each weight.  
\item Use an optimizer (e.g.\ Adam \cite{adam_ref}) that updates the weights $\bm{W}'_C = \bm{W}_C + \delta \bm{W}_C$ based on $\nabla_{\bm{W}_C} J$. 
\item Perform steps 2 to 6 in a loop until convergence is reached (as determined by a pre-selected convergence criterion).  
\end{enumerate}

In general, however, the total pulse duration is a function of $\phi$ \cite{fromonteil2024}. 
Further, physical errors -- arising from Doppler shifts due to the finite temperature of atoms, or decay from $\ket{r}$ -- propagate in time, and it is hence desirable to minimize the pulse duration. 

This can be included in our framework by adding a second NN: $N_T(\phi)$, with input $\phi$ and output $T_{\phi}$. We \emph{chain} $N_T$ and $N_C$ such that the output of $N_T$ is input into $N_C$, and denote this chained network by $N_{TC}$ -- see Fig.\ \ref{fig:network} for an example. To minimize the duration, the cost function is modified appropriately, 
\begin{equation}
\label{jopt}
J_{\mathrm{opt}} = J + \mu T_{\phi},
\end{equation}
for a multiplier constant $\mu$ that is selected in a heuristic manner, and the rest of the optimization proceeds as discussed above: with $\bm{v}_{\mathrm{in}}$ now input to $N_T$. 

As part of the steps outlined above, we must compute the gradient with respect to the NN weights  \mbox{$\bm{J}_{\mathrm{opt}}' = \nabla_{\bm{W}_C} J_{\mathrm{opt}}$}. The numerical solver of the Schr\"odinger equation can itself be interpreted as a NN, where the depth of the network is a continuous variable -- these are termed neural ODEs \cite{PhysRevLett.129.050507, autodiff2}. This identification allows us to backpropagate \cite{rumelhart_learning_1986} the cost function through the ODE solver: starting from $J_{\mathrm{opt}}$ we can repeatedly apply the chain rule backwards, first through the neural ODE, then the layers of $N_{TC}$, in order to compute $\bm{J}_{\mathrm{opt}}'$ \cite{autodiff, autodiff2}.
As the method is not specific to the Hamiltonian of the system, it is also generalizable to various quantum systems \cite{PhysRevLett.129.050507}. 

Randomly drawing $\bm{v}_{\mathrm{in}}$ in each iteration ensures that $J_{\mathrm{opt}}$ is evaluated over target angles previously unseen by the optimizer -- this ensures that $N_{TC}$, once converged, will return a high-fidelity pulse for all values of $\phi$.  

In practice, we find that splitting $\mathrm{dom}(\ckp)$ into multiple intervals and using a separate network $N_{TC}$ for each such interval leads to better solutions -- as quantified by a lower value for $J_{\mathrm{opt}}$. This results in control functions piecewise continuous in $\phi$. Operationally, these functions retain the advantage of the single $N_{TC}$ approach -- that outputs controls continuous in $\phi$ -- over standard QOC techniques. That is, they need to be trained only once to output high-fidelity pulses for any $\phi \in \mathrm{dom}(\ckp)$. Discussion of the justification and implications of this approach can be found in Appendix \ref{nnarch}.  

As an alternative approach, one could use QOC techniques like GRAPE to compute the pulse for multiple input angles in the domain, and interpolate between them to obtain a solution continuous in $\phi$ -- however, this interpolation step can introduce additional infidelities, and it fails if the pulses found at different angles belong to different families \cite{PhysRevLett.129.050507, lacroix_improving_2020}. We note recent work that adapts this interpolation approach to mitigate the associated infidelities \cite{dk2024}.

\section{Results}
\label{sec:results}
As discussed in Sec.~\ref{multiq}, to compute the infidelity of our pulses in a realistic experimental setup, we simulate decay from the Rydberg state, with estimated lifetime \mbox{$\tau = 96.5~\mathrm{\mu s}(\simeq 6\times 10^{3}/\Omega_{\mathrm{max}})$} for the $n = 61$ state \cite{Mohan_2023}.
Unless specified otherwise, a realistic blockade strength of $B = 21.1$ \cite{PhysRevResearch.4.033019} is also included in the simulation. 
We consider a Rabi frequency of $\Omega_{\mathrm{max}} = 2\pi \times 10 \mathrm{MHz}$ for Sr, which is within reach in the near term \cite{tsai2024, PhysRevResearch.4.033019}. 
Calculations of infidelities arising from the decay of $\ket{r}$, and from finite blockade, are discussed in Appendix \ref{appendix:block}. The total mean infidelity $\braket{(1 - F)}_{\phi}$ was computed over 200 samples randomly drawn from $\mathrm{dom}(\ckp)$.  
\begin{figure*}[htbp]
    \centering
    % First image
    \begin{subfigure}[b]{0.485\textwidth}
        \centering
        \includegraphics[width=\textwidth]{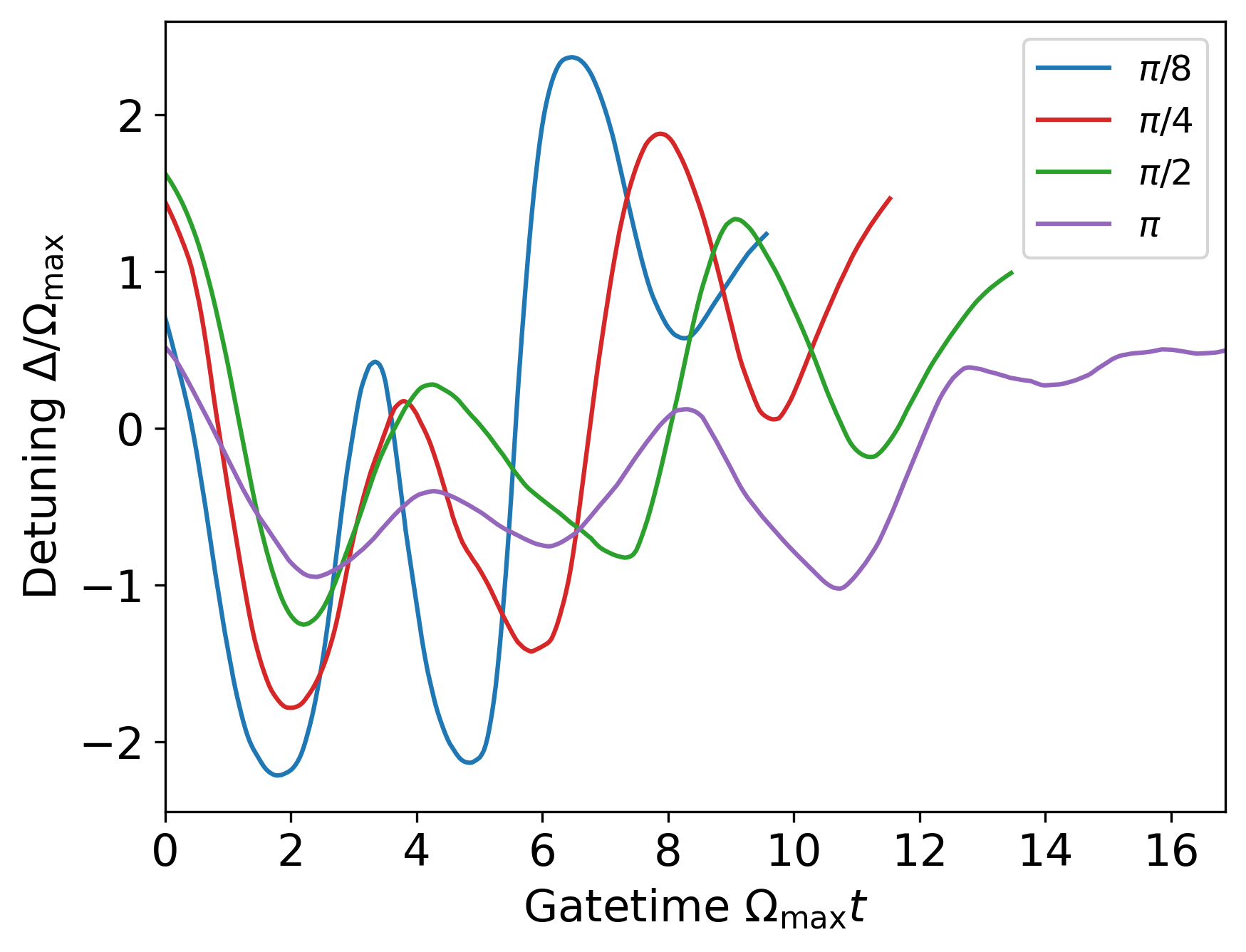}
        \caption{}
        \label{fig:pulses_3q}
    \end{subfigure}
    \hfill
    % Second image
    \begin{subfigure}[b]{0.49\textwidth}
        \centering
        \includegraphics[width=\textwidth]{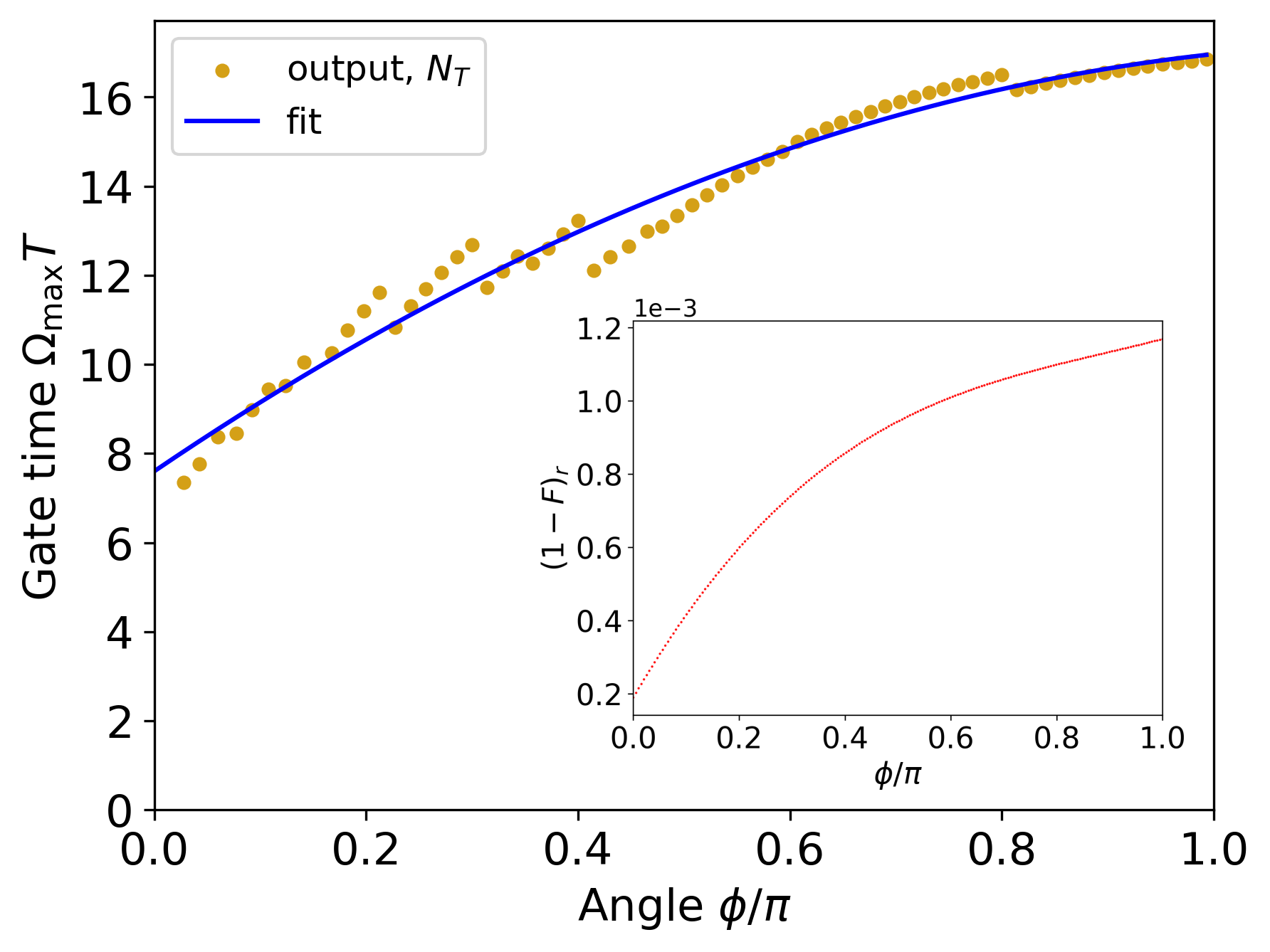}
        \caption{}
        \label{fig:time_3q}
    \end{subfigure}
    \captionsetup{justification=raggedright,singlelinecheck=false} % Only apply to this caption
    
    \caption{(a) Laser detuning $\Delta$ to implement the $\mathrm{C_2P}$ gate. (b) The $\mathrm{C_2P}$ pulse time (NN output, fit to second-order polynomial) as a function of $\phi$ and (inset) the infidelity arising from Rydberg state decay.}
    \label{fig:uni_3q}
\end{figure*}

As a consequence of using multiple networks for each gate, $T_{\phi}$ -- the output of $N_T$ -- is piecewise continuous in $\phi$. We will, however, fit the output of $N_T$ to continuous functions, in order to visualize and highlight the behavior of $T_{\phi}$. 
The fit constants were determined by a non-linear least squares approach.

The detuning range of the pulses is constrained to $|\Delta(t)|<2.5\Omega_{\mathrm{max}}$, as this simplifies the experimental implementation \cite{Mohan_2023}. See Appendix \ref{nnarch} for details about the constraint implementation. The optimized NNs are included as additional data \cite{additional_data}, and can be used to access the pulses for arbitrary $\phi$.

\subsection{$\mathrm{C_1P}$ gate}
\label{sec:c1pr}

Simulating the errors as discussed above, we obtain a infidelity $\braket{(1 - F)}_{\phi} = 3.4 \times 10^{-4}$ for the two-qubit $\mathrm{C_1P}$ gate over $\mathrm{dom(C_1P)}$ -- with the main error arising due to Rydberg state decay (see inset of Fig. \ref{fig:timec1p}). We also report the gate infidelity averaged over states in the Haar measure~\cite{nielsen2002}, $\braket{(1 -F')}_{\phi, \psi} = 2.7\times10^{-4}$. 

The pulses for some sample angles are plotted in Fig.~\ref{fig:uni}, along with $T_{\phi}$ as a function of $\phi$. We observe that the output from $N_T$ agrees closely with a fit to the function \begin{equation}
\label{fitfn}
f(\phi) = a\,\mathrm{arcsinh}(b\phi),
\end{equation}

for constants $a,b$, which is plotted in Fig. \ref{fig:timec1p}. This plot is consistent with currently known results for time-optimal $\mathrm{C_1P}$ pulses \cite{Evered_2023, fromonteil2024}. We consider finite $B$ in our optimization, compared to \mbox{$B\to\infty$} in the above studies, leading to minor differences ($\sim 2\%$ averaged over $\mathrm{dom(C_1P)}$) in pulse duration.

Here, we also plot the infidelities arising due to decay from $\ket{r}$ -- this indicates that not only $T_{\phi}$, but the total time spent in the Rydberg state also increases with $\phi$.  

\subsection{$\mathrm{C_2P}$ gate}
\label{sec:c2p}

We plot pulses for the $\mathrm{C_2P}$ gate, and the time-output of $N_T$ fit to a second-order polynomial, in Fig.~\ref{fig:uni_3q}. 

Considering only decay from the Rydberg state, we obtain an infidelity of \mbox{$\braket{(1 - F)}_{\phi} = 9.4 \times 10^{-4}$}. Including finite $B$, this jumps drastically to \mbox{$\braket{(1 - F)}_{\phi} = 1.54 \times 10^{-2}$}, thus rendering our pulses ineffective for the system we simulate. \par The error arising from finite blockade is well studied for two-qubit gates, and protocols often preserve a high fidelity despite such errors \cite{PhysRevA.94.032306, PhysRevA.101.062309, PRXQuantum.4.020335}. The three-qubit gates we obtain, however, have a greater pulse duration, and hence a lower fidelity for a finite $B$.

We hence study the strategy of using the $\mathrm{C_2P}$ NNs we trained at $B \to \infty$, as \emph{initial guesses} for $\mathrm{C_2P}$ NNs trained at $B = 21.1$, and then re-run the optimization as discussed in Sec. \ref{subsec:numop}. The pulses we obtain from this optimization reduce the average finite blockade infidelity to $\braket{(1-F)_{\mathrm{int}}}_{\phi} = 4.0 \times 10^{-4}$, and the total infidelity $\braket{(1 - F)}_{\phi} = 1.45 \times 10^{-3}$ -- where the decay from the Rydberg state is the main source of infidelity. The fidelity averaged over the states in the Haar measure, $\braket{(1-F')}_{\phi,\psi} = 1.28\times10^{-3}$. The results of this optimization are visualized in Fig. \ref{fig:blockade_3q}, in Appendix \ref{appendix:block}.

We note that we have to divide the domain in more steps and that hence more networks are needed than for the $\mathrm{C_1P}$ protocol -- $14$, compared to $5$ -- and the network architecture is more complex.
This leads to more pronounced discontinuities in $T_{\phi}$ with respect to $\phi$ for the $\mathrm{C_1P}$ pulses -- see Appendix \ref{nnarch}. 
 
Our approach hence converges to a high-fidelity solution, but not the time-optimal one (for which $T_{\phi}$ would be continuous in $\phi$, or agree closely with a continuous function as in Sec. \ref{sec:c1pr}). However, the pulse operates on a time scale comparable to the time-optimal solution -- the $\mathrm{C_2P(\pi)}$ pulse, for instance, has a duration of $T_{\pi} = 16.87/\Omega_{\mathrm{max}}$, which is around $2.6\%$ higher than the time-optimal $\mathrm{C_2Z}$ pulse \cite{Jandura_2022}.

%should we also 
\section{Conclusion}
\label{sec:discussion}

In this work we presented families of pulses that implement $\mathrm{C_1P}$ and $\mathrm{C_2P}$ gates on the neutral atom hardware with a high fidelity, in the presence of experimental sources of infidelity. Further, the NN weights we found could be valuable initial guesses for further optimizations -- for instance, to find pulses robust to hardware-specific sources of error like laser intensity noise \cite{Mohan_2023}.

It is of interest to extend these techniques for a larger number of control qubits, that is, $\ckp$ gates for $k>2$. NNs with a more complex architecture would be needed for such optimizations, and training such NNs without encountering vanishing gradients, or large runtimes, would benefit from additional machine learning techniques like regularization and batch normalization \cite{batchnormref}. The Hilbert space dimension $d = 3^{k+1}$ scales exponentially with the number of qubits, and will limit the $k$ achievable on a classical computer. However, we expect the training of NNs using graphics processing unit (GPU) clusters \cite{Iandola_2016_CVPR}, along with tensor network based methods for the simulation of the dynamics \cite{PhysRevLett.106.190501, PhysRevX.9.031041} to enable us to find controls for larger values of $k$. 

As discussed in the introduction, gate sets containing $\ckp$ gates can reduce circuit depth in quantum compilation, when compared to standard gate sets~\cite{staudacher2024}. Our results are of interest for this area of research, as they enable the realization of the $\ckp$-compiled circuits on neutral atom hardware. We are currently investigating resource-efficient state preparation schemes using these gates.

We note that a shorter version of the $\mathrm{C_2Z}$ gate -- exploiting an alternative definition of the $\mathrm{C_2Z}$ unitary -- has been proposed \cite{Evered_2023}, and an associated family of $\mathrm{C_2P}$ pulses may also be investigated. The corresponding optimization, however, requires an arbitrary global single-qubit rotation $U_1(\bm{\phi})$ instead of a correctional $R_Z$ rotation that one appends to the pulses, to implement the $\ckp$ unitary \cite{PhysRevResearch.4.033019}. This adds three functions (the rotation angles) to be optimized for, and is computationally expensive.

\section*{Data availability}
The weights for the NNs presented above, the parameters to specify their architecture (introduced in Appendix \ref{nnarch}), and a tutorial are included as additional data \cite{additional_data}. The code used for training the NNs is available upon reasonable request. 
\section*{Acknowledgements}
We acknowledge valuable discussions with Lucas Leclerc, Sven Jandura and Raul dos Santos. 
We note the use of the libraries PyTorch \cite{pytorch} and torchdiffeq \cite{torchdiffeq} for carrying out the NN-based optimization.

This research has received funding from the European Union’s Horizon 2020 research and innovation programme via the project 101070144 (EuRyQa) and under the Marie Skłodowska-Curie grant agreement number 955479. S.K.\ acknowledges funding from the Dutch Ministry of Economic Affairs and Climate Policy (EZK) as part of the Quantum Delta NL programme, and from the Netherlands Organisation for Scientific Research (NWO) under Grant No.\ 680.92.18.05.
\appendix
\section{NN architecture and training}
\label{nnarch}

Consider a feedforward NN consisting of $m_L$ layers, each with $m_N$ neurons (with the exception of the input and output layers) -- we visualized two such networks in Fig. \ref{fig:network}, both with $(m_L, m_N) = (4, 3)$. 
We denote the weights in each layer $l$ by $\bm{W}_l$ -- each weight is represented by an edge in Fig. \ref{fig:network}, and hence $\bm{W}_l$ is a matrix. 
The network $N_T$ and $N_C$ in Fig. \ref{fig:network}, for instance, consist of $4$ layers with $3$ neurons in each intermediate, or \emph{hidden}, layer.

Each neuron itself represents a weighted sum of the inputs (along with a \emph{bias term} $\bm{b}_l$), as well as an \emph{activation function}. The output from the $n^{\mathrm{th}}$ layer is 
\begin{equation}
\label{forward}
\bm{x}_n = f(\bm{W}_n\bm{x}_{n-1} + \bm{b}_n),
\end{equation}
for \emph{activation function} $f$ that introduces non-linearities into the network.  We use \emph{fully connected} NNs in our research -- where a neuron in a layer $k$ is connected, by an edge, to every neuron in layer $k+1$.  

We choose the sigmoid activation function \mbox{$f_{\sigma}(y) = 1/(1 + e^{-y})$} for the output layers, as the output can be conveniently bounded to some desired value -- we bound $|\Delta(t)| < 2.5\Omega_{\mathrm{max}}$ to make the pulse implementation more feasible on current generation hardware. We bound $T_{\phi} < 1.2T_{\mathrm{opt}, \phi = \pi}$, i.e. by the optimal time for the $\mathrm{C_1Z}$ and $\mathrm{C_2Z}$ gates, which are known \cite{Jandura_2022} -- slightly higher than $T_{\mathrm{opt}, \phi = \pi}$ as we wish to avoid vanishing gradients which arise for $f_{\sigma} \simeq 1$. This restricts the solution space and hence aids convergence to the true solution. 
We choose the \emph{rectified linear unit} (ReLU) activation function for the intermediate layers, 
$f_{\mathrm{ReLU}}(x) = \max (0, x)$. 

The NN architecture for $N_{TC}$ is determined by the choice of the activation function, and by $(m_L^T, m_N^T, m_L^C, m_N^C)$ -- the layers and intermediate layer neurons for $N_T$ and $N_C$ respectively. The deeper and wider the network (the larger the $m_L$ and $m_N$ respectively), the larger the variational space -- hence the higher the likelihood that the true solution lies in the space.
However, this complexity can also make it difficult to train the network -- such large NNs have the tendency to suffer from vanishing gradients \cite{vanish}. 
\par A balance must hence be struck when choosing $m_L$ and $m_N$ to account for both the above factors -- we note for instance that the optimization for the $\mathrm{C_2P}$ pulses is more challenging than for $\mathrm{C_1P}$, owing in part to the larger dimension of the Hamiltonian ($27\times27$ as opposed to $9\times 9$), more phases (picked up by the different computational states) to account for \cite{Jandura_2022}, and a longer pulse time. 

We choose these parameters in a heuristic manner. In principle, one $N_{TC}$ with the appropriate architecture should be able to model the pulses over $\mathrm{dom}(\ckp)$. As discussed above, however, finding this architecture is a challenge. On using deep NNs, we often encountered vanishing gradients that slow down convergence -- this occurs for $m^C_L > 20, m^C_N > 300$, the parameters for $N_C$.  

As presented in the main text, an alternative approach is to use NNs with fewer layers and weights, that can accurately model the pulses over a part of the $\mathrm{dom}(\ckp)$. We then split this domain into multiple parts and assign one NN to be trained for each part. 

For the $\mathrm{C_1P}$ NNs we used the architecture $(3, 45, 10, 300)$, and for the $\mathrm{C_2P}$ NNs $(4, 45, 20, 300)$. 

This approach could potentially allow for multiple families of pulses to be found. Operationally, however, this does not make a difference compared to using one $N_{TC}$ -- both approaches return pulses that implement the desired gates over the entire domain. Further, we use the converged NN trained over one part of $\mathrm{dom}(\ckp)$ as the initial guess for all other parts, to speed up convergence -- this also reduces the likelihood of obtaining a separate pulse family. 

Using the multiple $N_{TC}$ approach also implies that discontinuities might arise in $T_{\phi}$ at the \emph{junction points} of the domain parts -- that is, $T_{\phi}$ is piecewise continuous instead of continuous. For the $\mathrm{C_1P}$ these discontinuities are minor and we obtain a good fit for $T_{\phi}$ with Eq.~(\ref{fitfn}), and with results from other studies \cite{Evered_2023, fromonteil2024}, as discussed in the main text. For the $\mathrm{C_2P}$ optimization these discontinuities are more significant, partly because more $N_{TC}$ were needed -- $14$ as opposed to $5$ for $\mathrm{C_1P}$ -- introducing more junction points. 

The constants for the function in Eq.~(\ref{fitfn}), determined by fitting to the output of $N_T$ for the $\mathrm{C_1P}$ protocol are $(a, b) = (1.07, 275.86)$. Similarly, the constants for the second-order polynomial $f(x) = ax^2+bx+c$ for the $\mathrm{C_2P}$ protocol are $(a, b, c) = (-0.70, 5.24, 7.44)$.   

\section{Losses from finite blockade and Rydberg state decay}
\label{appendix:block}
\begin{figure}
        \centering
\hspace*{-2cm}
\includegraphics[width=0.55\textwidth]{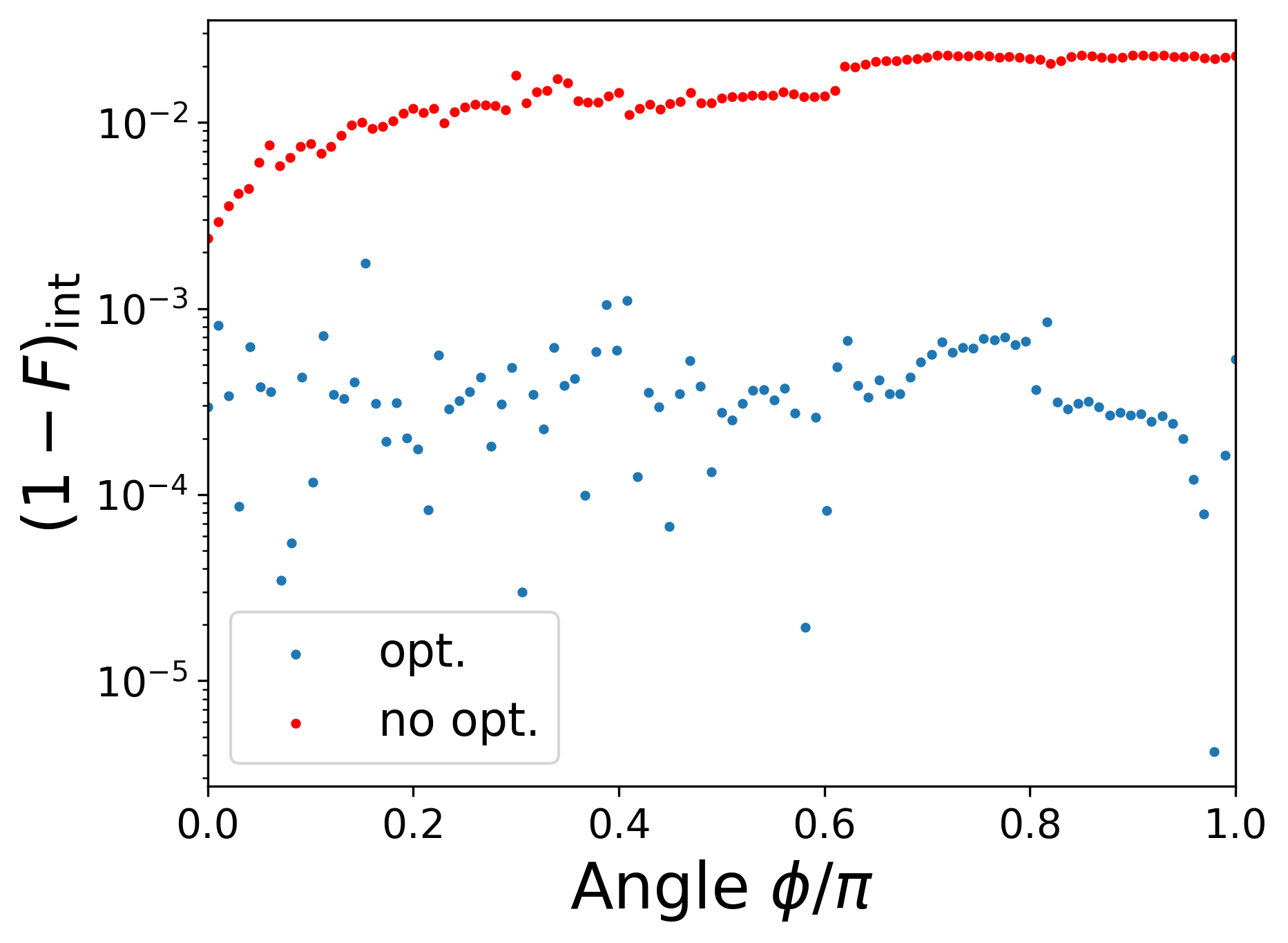}
\captionsetup{justification=raggedright,singlelinecheck=false} 
        \caption{Infidelities arising from finite blockade strengths for unoptimized \emph{infinite blockade} pulses (red) and optimized pulses (blue). Our method is successfully able to mitigate the blockade-induced losses.}
        \label{fig:blockade_3q}
\end{figure}
As discussed in Sec.\ \ref{multiq}, the assumption of blockade strength \mbox{$B\to\infty$} 
implies that states with more than one Rydberg excitation cannot be populated during the pulse. This allows us to project them out by defining an effective Hamiltonian $H_{\mathrm{eff}}$, in order to reduce the computational cost of the simulation, with the potential downside of added infidelities in a realistic system (with finite $B$), as we explored in Sec.\ \ref{sec:c2p}. 

First, for some operator $G$, we define $P_{\Sigma}(G)$ as an operator constructed from the sum over \emph{distinct permutations} of qubit states, for instance,
\begin{align}
P_{\Sigma}(\ket{011}(\bra{0}_1\otimes\bra{b_2}_{2,3})) = \ket{011}(\bra{0}_1\otimes\bra{b_2}_{2,3})\nonumber\\ + \ket{101}(\bra{0}_2\otimes\bra{b_2}_{1,3}) + \ket{110}(\bra{0}_3\otimes\bra{b_2}_{1,2}).
\end{align}

For the $\mathrm{C_2P}$ protocol over three qubits, the Hamiltonian in the limit $B\to\infty$ is:
\begin{align}
\label{eq:h_eff}
H_{\mathrm{eff}} &= \Big[\frac{\Omega(t)}{2}P_{\Sigma}(\ket{001}\bra{00r}) \nonumber\\&+\frac{\sqrt{2}\Omega(t)}{2}P_{\Sigma}(\ket{011}(\bra{0}_1\otimes\bra{b_2}_{2,3})) \nonumber\\&+ \frac{\sqrt{3}\Omega(t)}{2}\ket{111}\bra{b_3} \nonumber\\ 
&- \Delta(t)\sum_{i,j\in\{0,1\}}P_{\Sigma}(\ket{ijr}\bra{ijr})\Big] + h.c.
\end{align}

The infidelity due to decay from $\ket{r}$ was defined as 
\begin{equation}
\label{eq:spem}
(1-F)_r = F - F_{\mathrm{decay}},
\end{equation}
for $F, F_{\mathrm{decay}}$ the fidelities are obtained by simulating the pulses under $H_{Nq}$ [see Eq.~(\ref{1rtrans})], and $H_{Nq,\Gamma}$ respectively.     
The infidelity arising from finite blockade was defined for three qubits, in a manner similar to Eq.~(\ref{eq:spem}), \mbox{$(1-F)_{\mathrm{int}} = F_{\infty} - F_{\mathrm{fin}}$}, for $F_{\infty}, F_{\mathrm{fin}}$ the fidelities of the pulses simulated for $H_{\mathrm{eff}}$ and $H_{3q}$ respectively. This infidelity is visualized as a function of $\phi$ in Fig. \ref{fig:blockade_3q}.

We expect that the pulses optimized for some constant $B$ will not retain a high fidelity for \mbox{$B\to\infty$}, or for some $B'$ that differs significantly from the value considered. To obtain pulses that have a high fidelity at $B'$, the optimization has to be carried out again -- the NN weights provided in the additional data serve as good initial guesses for this optimization.  

\section{Native gates and platforms}
\label{appendix:decomp}
As discussed in Sec. \ref{multiq}, it is possible to implement the $\ckp$ unitary by expressing it as a product of hardware-available gate unitaries -- we term this a decomposition based approach. An example of this approach, expressing the $\mathrm{C_1P}$ unitary in terms of single-qubit gates and the $\mathrm{C_1Z}$ gate, was presented in Fig. \ref{fig:native}. The native approach, which we looked at, implements the $\ckp$ unitary directly on the pulse level -- that is, it is implemented by the dynamics induced by the laser pulses on the electronic states of the atoms. 

It is worth comparing these two approaches for NISQ-era quantum computers. Qualitatively, finding optimal decompositions is a difficult problem, and it is complicated further if we include $\ckz$ gates in the standard gate set, as is the case for neutral atoms. We hence consider decompositions of the unitaries in terms of one- and two-qubit gates. 

Let us define $T_{\mathrm{d}, k}$ as the total time taken to implement two-qubit gates in the decomposition approach for $\ckp$, and $T_{\mathrm{n}, k}$ as the time the native approach takes, averaged across the domain. We then propose the ratio $R_k = T_{\mathrm{d}, k}/T_{\mathrm{n}, k}$  as a metric for comparison. As one-qubit gates can be implemented with significantly higher fidelities \cite{PhysRevLett.121.240501, universal2024}, we do not include them in $T_{\mathrm{d}, k}$.

For the $\mathrm{C_1P}$ unitary, using the decomposition of Fig. \ref{fig:native} and time-optimal $\mathrm{C_1Z}$ gates \cite{Jandura_2022}, we obtain $R_1 = 2.2$. The $\mathrm{C_2P}$ gate can be decomposed into a product of two $\mathrm{CNOT}$ gates and three $\mathrm{C_1P}$ gates \cite{ccpdecomp}, which is equivalent to $8$ $\mathrm{C_1Z}$ gates, and $R_2 = 4.6$. It is expected that $R_k$ increases with $k$ \cite{PhysRevLett.129.050507}. 

We note that this metric will differ depending on the decomposition scheme -- it would be of interest to devise such schemes that account for the availability of $\ckz$ gates in the standard gate set. 

It is also worth highlighting the native $N$-qubit gate protocols (for $N>2$) that have been proposed or demonstrated on various quantum platforms. This includes native implementations of $N$-qubit Toffoli gates~\cite{PhysRevA.103.052437} and M\o lmer-S\o rensen entanglement schemes~\cite{PhysRevLett.82.1835} on trapped ion systems. Further, $\mathrm{C_2P}$ protocols have also been implemented on the superconducting platform~\cite{PhysRevApplied.19.044001}.

In terms of gates, it is straightforward to devise native $N$-qubit protocols on the neutral atom platform. This is in part due to the long range interactions, and partly due to \emph{reconfigurable} arrays--the ability to move the atoms around the tweezer during computation, allowing for gates over distant qubits without additional lossy SWAP operations~\cite{Bluvstein_2023}. As noted in the Introduction, gates on up to $9$ qubits have been experimentally demonstrated~\cite{cao2024multiqubit}, and several protocols have been proposed for $\mathrm{C}_k\mathrm{Z}$ gates with an arbitrary number of controls $k$.

Trapped ion systems enjoy similar benefits to multiqubit gate implementations.   
In contrast, such gates on superconducting quantum computers currently require significant modifications to the hardware and are difficult to scale to multiple control qubits~\cite{PhysRevApplied.19.044001}. Recent studies also demonstrate native multiqubit gates with quantum dots~\cite{cao2019} and solid-state systems~\cite{wang2023}, and hence a systematic comparison of these platforms is a timely question, made challenging by their rapid pace of development.
\bibliography{mainbib}

\end{document}